\documentclass{JHEP3}

\newcommand{\cN}{{\mathcal N}}

\usepackage{latexsym,amsfonts,amsmath,amssymb,graphics,amsthm,eucal,graphicx}
\usepackage{multirow,epsfig,amsmath}

\title{$\mathcal{N}=2$ Dualities and $Z$-extremization in Three Dimensions}
\author{Brian Willett\\ California Institute of Technology \\ Email: \email{bwillett@caltech.edu}}
\author{Itamar Yaakov\\ California Institute of Technology \\ Email: \email{itamar.yaakov@caltech.edu}}

\abstract{We use localization techniques to study duality in $\mathcal{N}=2$ supersymmetric gauge theories in three dimensions. Specifically, we consider a duality due to Aharony involving unitary and symplectic gauge groups, which is similar to Seiberg duality in four dimensions, as well as related dualities involving Chern-Simons terms. These theories have the possibility of non trivial anomalous dimensions for the chiral multiplets and were previously difficult to examine. We use a matrix model to compute the partition functions on both sides of the duality, deformed by real mass and FI terms. The results provide strong evidence for the validity of the proposed dualities. We also comment on a recent proposal for recovering the exact IR conformal dimensions in such theories using localization.}
\preprint{}

\begin{document}

\section{Introduction}

Supersymmetric gauge theories in three dimensions may have multiple effective descriptions of their IR dynamics. One example of such an IR duality is mirror symmetry of $\cN=4$ quiver gauge theories \cite{Intriligator:1996ex}. Other examples include the dualities for $\cN\ge 3$ Chern Simons gauge theories proposed in \cite{Giveon:2008zn} and \cite{Aharony:2008gk}. The large extended supersymmetry and non-Abelian R-symmetry present in these theories implies a vanishing anomalous dimension for the chiral matter multiplets. There exist $\cN=2$ versions of both mirror symmetry \cite{deBoer:1996ck}\cite{deBoer:1996mp}\cite{Aharony:1997bx} and Seiberg-like dualities \cite{Aharony:1997gp}\cite{Giveon:2008zn}. Theories with $\cN=2$ supersymmetry in three dimensions, which corresponds to $\cN=1$ in four dimensions, are much richer, allowing for an arbitrary superpotential and anomalous dimensions for chiral fields. Such theories still have holomorphy properties that enable us to do some calculations exactly. $\cN=2$ theories may possess a $U(1)$ R-symmetry, the automorphism group of the $\cN=2$ supersymmetry algebra. The Noether current for this R-symmetry generically mixes with the currents for other $U(1)$ global symmetries as we flow to the IR. These additional symmetries may include flavor symmetries manifest in the Lagrangian, the topological $U(1)_J$ symmetry, with current $\star F$, as well as possible hidden symmetries. At the IR fixed point, a distinguished combination of such conserved currents, the IR R-charge, sits in the same supermultiplet as the energy momentum tensor. This restricts the conformal dimension of all operators to be no less than their R-charge. The inequality is saturated for chiral operators. 

Using localization techniques, the path integral calculation for a supersymmetric observable can sometimes be reduced to a finite dimensional integral \cite{Pestun:2009nn}. In previous work, such a reduction was performed for three dimensional superconformal gauge theories \cite{Kapustin:2009kz}. The resulting matrix model can be used to compute the partition function of a wide class of theories, such as the recently introduced superconformal Chern Simons matter theories, and the IR fixed points of gauge theories with Yang Mills terms. The calculation involves a conformal transformation to $S^3$, and depends, crucially, on knowing the IR conformal dimensions of all the fields. In previous checks of IR duality, it was implicitly assumed that the fields have canonical scaling dimension \cite{Kapustin:2010xq}.  Recently, this assumption was relaxed and a matrix model was derived for theories with matter of arbitrary dimension \cite{Jafferis:2010un}.  In this paper we use this generalized matrix model to test some dualities that were beyond the reach of the original matrix model.

We will compare the partition functions for $\cN=2$ gauge theories discussed in \cite{Aharony:1997gp} and \cite{Giveon:2008zn}. One can deform these theories, in a supersymmetric manner, by weakly gauging any of the global $U(1)$ symmetries and giving an expectation value to the scalar in the background vector or linear multiplet.  This has the effect of giving each of the fields a real mass proportional to its charge under the symmetry, or, in the case of the topological $U(1)_J$ symmetry, an FI term.  If one performs this operation on two theories related by a duality, and if the relevant symmetries are mapped to each other under the duality, the partition functions should agree as a function of the deformations.  This provides a more robust check of the duality than the matching of the partition functions alone.

The supersymmetric deformations are closely related to the ambiguity of the IR $R$-symmetry, as follows. The possible $R$-symmetries of a theory can be shown to differ by an Abelian global symmetry. In \cite{Jafferis:2010un}, it was shown how to compute the partition function $Z$ for a given trial $R$-symmetry. It was also argued that the correct $R$-symmetry is the one that extremizes $|Z|$. For a given Abelian symmetry, it was shown that the partition function is holomorphic in the combination $m+iq$, where $m$ is the expectation value of the scalar in the background vector or linear multiplet used to weakly gauge the symmetry, as above, and $q$ is the contribution of the associated current to the IR $R$-symmetry. It follows that if the partition functions agree as holomorphic functions of the mass deformations, they also agree as one varies the trial $R$-symmetry in the appropriate way on both sides. This means that one does not need to know the correct IR $R$-symmetry to test these dualities\footnote{We would like to thank David Kutasov for explaining this point to us.}. One simply needs to understand how the global symmetries map.  Knowing the correct IR $R$ symmetry would be equivalent to knowing the origin of this space of deformations.  On the other hand, this also means one cannot use the duality to determine the correct IR $R$-symmetry.  One needs other methods to do this, such as extremizing the value of $|Z|$  \cite{Jafferis:2010un}, which we briefly discuss in the last section.

\acknowledgments{We would like to thank Anton Kapustin, David Kutasov, and Fokko van de Bult for useful discussions while completing this paper.}

\section{Localization}

In this section we describe the localization procedure used in calculating the partition functions of gauge theories in three dimensions. A more detailed explanation of the deformation used to localize the action, and the derivation of the resulting matrix model, can be found in \cite{Kapustin:2009kz} and \cite{Kapustin:2010xq}. The generalization to chiral multiplets of arbitrary conformal dimension is found in \cite{Jafferis:2010un}.

We consider the superconformal field theory which is the IR fixed point of a supersymmetric gauge theory. We will consider both theories with and without Chern Simons terms. After a conformal transformation to $S^3$ the action is deformed by a $Q$ exact term, where $Q$ is a particular fermionic generator in the supersymmetry algebra.  In the limit where the deformation is very large, the path integral localizes to a finite dimensional subspace parameterized by a single matrix in the adjoint of the gauge group. The remaining integral is over this matrix or, equivalently, over its eigenvalues. The ingredients of the resulting matrix model were given in \cite{Kapustin:2009kz}. We describe only the relevant components.

A gauge field coupled to charged chiral multiplets must have conformal dimension $1$ in the IR. This can be deduced by considering the topological current $\star F$ which is conserved and is therefore of conformal dimension $2$. If the gauge field is free, it may be dualized to a free scalar and would have conformal dimension $3/2$. In this case, the current $\star F$ is not a conformal primary. We assume that this does not happen for any of the theories in questions, so that the contribution of the gauge sector does not change from \cite{Kapustin:2009kz}, and is given by:

\[ Z_{1-loop}^{gauge} ( \sigma) = \det_{Ad} \frac{2 \sinh (\pi \sigma)}{\pi \sigma} \]

We may pass to the Cartan of the gauge group $G$, parameterized by the eigenvalues $\lambda_j$, $j=1,...,\mbox{Rank}(G)$.  Then the Vandermonde determinant cancels against the denominator of the above expression, and the resulting determinant can be written as a product over the roots of the Lie algebra.  In this paper we will consider the groups $U(N)$ and $Sp(2N)$, both of rank $N$, and the corresponding $1$-loop determinants are given by, for $U(N)$:

\begin{equation}
\label{unfac}
\prod_{1 \leq i<j \leq N} (2 \sinh \pi (\lambda_i - \lambda_j))^2
\end{equation}
and, for $Sp(2N)$:

\begin{equation}
\label{spnfac}
\prod_{1 \leq i < j \leq N} \bigg( (2 \sinh \pi (\lambda_i - \lambda_j))^2 (2 \sinh \pi (\lambda_i + \lambda_j))^2 \bigg) \prod_{j=1}^N (2 \sinh (2 \pi \lambda_i))^2
\end{equation}

All gauge multiplets will have the conventional Yang-Mills kinetic term in the UV.  In addition, there may be a Chern-Simons term at level $k$, whose contribution is:

\[ \prod_{j=1}^N e^{-k \pi i {\lambda_j}^2} \]

\subsection{Matter}

Next we consider the contribution of chiral multiplets.  The results reviewed in this section can be found in \cite{Jafferis:2010un}.

In general, a chiral multiplet comes in a certain representation of both the gauge group and the global flavor symmetry group of the theory.  These can be treated somewhat symmetrically by weakly gauging the flavor symmetries, which can be seen as follows.  As described above, for each gauge field there is a scalar partner $\sigma$, and the matrix model is an integral over its zero modes.  If we have a background gauge field, one can also consider giving an expectation value to the corresponding scalar, $\sigma_{BG}$, and it will enter the matrix model in the basically same way as a dynamical $\sigma$.  The only difference is that we do not integrate over the background $\sigma_{BG}$, rather it is a parameter that we can tune.  After reducing the integral to one over the Cartan, parameterized by the eigenvalues $\lambda_j$ of $\sigma$, the eigenvalues for the background vector multiplets correspond to real masses for the fields.

Now consider a chiral multiplet whose fields have canonical dimension, ie, the scalar has dimension $\frac{1}{2}$.  After we reduce the gauge and global symmetry groups to their maximal torii, we can list the charges $q_a$ of this multiplet under each $U(1)$ factor.  Then, if $\lambda_a$ denotes the corresponding eigenvalue, the $1$-loop determinant is given by \cite{Jafferis:2010un}:

\[ e^{\ell(\frac{1}{2} + i \sum_a q_a \lambda_a)} \]
where:\footnote{See the appendix for more discussion on this function.}

\[ \ell(z) = - z \log ( 1 - e^{2 \pi i z} ) + \frac{i}{2} ( \pi z^2 + \frac{1}{\pi} \mbox{Li}_2(e^{2 \pi i z})) - \frac{i \pi}{12} \]

For theories with at least $\mathcal{N}=3$ supersymmetry, the chiral multiplets are grouped into hypermultiplets, pairs of chiral multiplets in conjugate representations.  In addition, the non-abelian $R$-symmetry protects the fields from corrections to the dimension as we flow to the IR.  Thus the contribution of a hypermultiplet is:

\[ e^{\ell(\frac{1}{2} + i \sum_a q_a \lambda_a) + \ell(\frac{1}{2} - i \sum_a q_a \lambda_a)} = \frac{1}{2 \cosh \pi(\sum_a q_a \lambda_a)} \]

For theories with only $\mathcal{N}=2$ supersymmetry, things are more complicated.  Now the $R$-symmetry is abelian, and we do not get the same non-renormalization theorem we had before.  As argued in \cite{Jafferis:2010un}, the contribution of a chiral multiplet of dimension\footnote{We will define the dimension of a chiral multiplet to be the dimension of its dynamical scalar.  In particular, canonical dimension corresponds to dimension $\frac{1}{2}$.} $\Delta$ is given by:

\[ e^{\ell(1 -\Delta + i \sum_a q_a \lambda_a)} \]

Let us elaborate on this.  In the UV, the theory is not conformal, so there is no privileged $R$-symmetry.  Any abelian symmetry that does not commute with the supersymmetry generators will do, and any two of these will differ by an abelian flavor symmetry.  It will be convenient to use this freedom to set the UV $R$-charge of many of the fields to be $\frac{1}{2}$, and we will make this choice when possible, calling the result ``the'' UV $R$-charge.

At the IR fixed point, there is a unique choice of $R$-symmetry whose current lies in the same multiplet as the stress-energy tensor.  We can write it as:

\[ R_{IR} = R_{UV} + \sum_a c_a Q_a \]
where $Q_a$ runs over the abelian global symmetries of the theory.  In a superconformal field theory, the $R$ charge and scaling dimension of a chiral primary are the same, as a consequence of the superconformal algebra.  This means we can write the dimension of the chiral as:\footnote{Here we have assumed the UV $R$-charge is $\frac{1}{2}$.  We will encounter a few exceptions later on, and the appropriate modification will be made.}

\[ \Delta = \frac{1}{2} + \sum_a c_a q_a \]
and the $1$-loop determinant becomes:\footnote{For simplicity we let the index $a$ run over all symmetries of the theory, although for gauge symmetries there is no contribution to the $R$-symmetry, and so the corresponding $c_a$ are zero.}

\[ e^{\ell(\frac{1}{2} + i \sum_a q_a(\lambda_a + i c_a))} \]

In other words, shifting the $R$-symmetry by a flavor symmetry is equivalent to weakly gauging that symmetry and giving the background scalar a \textit{complex} value.

As an example, for a hypermultiplet of canonical dimension in a fundamental representation of $U(N)$, the $1$-loop partition function is given by:

\[ \prod_{j=1}^N e^{\ell(\frac{1}{2} + i \lambda_j) + \ell(\frac{1}{2} - i \lambda_j)} \]
We can now consider giving different masses $m$ and $\tilde{m}$ to the two chirals:

\[ \prod_{j=1}^N e^{\ell(\frac{1}{2} + i \lambda_j + i m) + \ell(\frac{1}{2} - i \lambda_j - i \tilde{m})} \]
When the masses are the same for the two chirals, we call this a vector mass for the hypermultiplet, while if they differ by a sign, we call it an axial mass.  Giving these masses complex values corresponds to mixing the $R$-symmetry with the $U(1)$ symmetries rotating the phases of these chiral multiplets.  Typically there is a symmetry exchanging the two chiral multiplets which forbids the vector mass from contributing, but the axial mass parameter will in general be complex.

We close the section by mentioning that, in addition to flavor symmetries manifest in the lagrangian, the $R$-symmetry can mix with other, more subtle global symmetries.  This includes any topological $U(1)_J$ symmetries, whose current $\star F$ is conserved by virtue of the Bianchi identity, as well as hidden symmetries that appear at the IR fixed point, but are not visible in the UV description of the theory.  For the former, we can still gauge the symmetry, and the corresponding deformation corresponds not to another real mass term, but instead to an FI term $\eta$, which enters the matrix model by an insertion of:

\[ e^{2 \pi i \eta \sum_j \lambda_j} \]
In analogy to what we did with the flavor symmetries, one can allow for the possibility that the $R$-symmetry mixes with this symmetry by letting $\eta$ become complex.  We will not have much to say about hidden symmetries at this point, although we will find that they probably do arise and play an important role in many of the theories we will consider.

\section{Aharony-Seiberg Duality}

In the next two sections we will test a few proposals for dualities between $\mathcal{N}=2$ gauge theories.  These theories all have conventional Yang-Mills terms for the gauge field in the UV, in addition to Chern-Simons terms in the examples of the next section.  In three dimensions, the gauge coupling is dimensionful, and so none of these theories are conformal.  Thus the duality is between their IR fixed points, which are generically strongly interacting theories.  One is able to provide evidence for these dualities by using the matrix model to compute the partition function of these strongly coupled superconformal theories.  This was done in earlier papers for theories with at least $\mathcal{N}=3$ supersymmetry \cite{Kapustin:2010xq} \cite{Kapustin:2010mh}. 

As described above, in addition to testing the mapping of the partition functions, one can deform them by weakly gauging the flavor symmetries to add real masses or FI terms, and showing the partition functions agree as a function of these deformations.  This provides evidence not only for the duality, but also for the proposed mapping of global symmetries between the two theories.

For theories with $\mathcal{N}=2$ supersymmetry, the dimensions of the fields in the IR are unknown, as the $R$ symmetry may mix arbitrarily with abelian flavor symmetries.  As described in the previous section, one can account for this by allowing the mass deformations to become complex.  Varying the $R$-symmetry corresponds to varying the imaginary parts of these mass parameters, and in principle there is one choice which is correct.  One might be concerned that it is impossible to check the duality without knowing the correct IR $R$-symmetry.  However, it will turn out that the partition functions agree as analytic functions of the mass deformations, so it is unnecessary to know the correct IR $R$-symmetry: the duality works for any possible $R$-symmetry.  We will describe this in more detail in the examples below.

In the present section, we consider two classes of dualities studied by Aharony in \cite{Aharony:1997gp}.  These are reminiscent of Seiberg duality in four dimensions, so we will call this Aharony-Seiberg duality.  In that paper, the main evidence presented for the dualities were the matching of the moduli spaces.  In order to achieve this matching, certain singlet chiral fields need to be added to the dual theory, parameterizing the Coulomb branch of the original theory, and a superpotential coupling this field to a monopole operator must be included.  We will find that it is necessary to include the $1$-loop partition functions for these extra fields in order to achieve precise matching of the partition functions, although this test is not sensitive to the form of the superpotential.

\subsection{Unitary Group}

The first duality involves two $\mathcal{N}=2$ gauge theories with a unitary gauge group.  The dual theories are \cite{Aharony:1997gp}:

\begin{itemize}
\item $\mathcal{N}=2$ $U(N_c)$ gauge theory with $N_f$ fundamental chiral multiplets $Q_a$ and $N_f$ anti-fundamental chiral multiplets $\tilde{Q}^a$.  We will call a single pair $(Q_a,\tilde{Q}^a)$ a flavor.  There is no superpotential.
\item $\mathcal{N}=2$ $U(N_f-N_c)$ gauge theory with $N_f$ fundamental flavors.  In addition, there are ${N_f}^2$ uncharged chiral multiplets ${M_a}^b$ and two uncharged chiral multiplets $V_\pm$, which couple via the following superpotential:

\[ \tilde{q}_a {M^a}_b q^b + V_+ \tilde{V}_- + V_- \tilde{V}_+ \]

where $\tilde{V}_\pm$ are monopole operators, parameterizing the Coulomb branch of this theory.
\end{itemize}

Note that $V_\pm$ are fundamental (ie, non-composite) fields, while $\tilde{V}_\pm$ are monopole operators, so can in principle be expressed in terms of the other fields.  In fact, $V_\pm$ are mapped under the duality to the monopole operators of the first theory, while ${M^a}_b$ is mapped to $Q^a \tilde{Q}_b$.

Now let us discuss the flavor symmetries of these two theories, and how they are mapped under the duality.  For both theories, there is in principle a $U(N_f) \times U(N_f)$ flavor symmetry rotating the two sets of chiral fields.  However, the diagonal $U(1)$ is gauged, so this is reduced to $SU(N_f)\times SU(N_f) \times U(1)_A$.  In addition, there is a $U(1)_J$ topological symmetry, whose current is $\star \mbox{Tr} F$.  The $V_\pm$ fields are charged under both $U(1)_A$ and $U(1)_J$.

Note that the symmetry group is the same for both theories. This means one can summarize how the duality acts on these symmetries by thinking of a single symmetry group which acts on both theories, and listing the charges of the fields of both theories under this group.  We summarize this in the following table:

\[\]
\begin{center}
\begin{tabular}{| c | c  c  c c | } 
\hline
Field & $SU(N_f) \times SU(N_f)$ & $U(1)_A$ & $U(1)_J$ & $U(1)_{R-UV}$ \\
\hline
$Q_a$ & $(N_f,1)$ & $1$ & $0$ & $\frac{1}{2}$ \\
$\tilde{Q}^a$ & $(1,\bar{N}_f)$ & $1$ & $0$ & $\frac{1}{2}$\\
\hline
$q^a$ & $(\bar{N}_f,1)$ & $-1$ & $0$ & $\frac{1}{2}$\\
$\tilde{q}_a$ & $(1,N_f)$ & $-1$ & $0$ & $\frac{1}{2}$\\
${M^a}_b$ & $(N_f,\bar{N}_f)$ & $2$ & $0$ & $1$\\
$V_\pm$ & $(1,1)$ & $-N_f$ & $\pm 1$ & $\frac{N_f}{2}-N_c+1$\\
\hline
\end{tabular}
\end{center}
\[\]

Corresponding to the two $SU(N_f)$ factors, we add masses for the two chiral multiplets in each flavor, $m_a$ and $\tilde{m}_a$, which are each constrained to sum to zero.  In addition, for $U(1)_A$ there is an total axial mass $\mu$, and for $U(1)_J$ there is the FI term $\eta$.  Including all of these deformations, the partition function for the first theory can be written as:

\[ Z_{N_f,N_c}^{(U)}(\eta;m_a;\tilde{m}_a;\mu) = \frac{1}{N_c!} \int \prod_{j=1}^{N_c} \bigg( d \lambda_j \prod_{a=1}^{N_f} e^{\ell(\frac{1}{2} + i \lambda_j + i m_a + i \mu)+\ell(\frac{1}{2} - i \lambda_j - i \tilde{m_a} + i \mu)} \bigg) \prod_{i<j} (2 \sinh \pi (\lambda_i - \lambda_j))^2 \]

For the second theory, we see that the representation of $SU(N_f) \times SU(N_f) \times U(1)_A$ in which the quarks lie is replaced by its conjugate, so all mass terms should come in with the opposite sign.  Inspecting the table above, we see that the $1$-loop partition function for ${M^a}_b$ is:

\[ e^{\ell( i(m_a - \tilde{m}_b + 2 \mu)} \]
while that of $V_\pm$ is:

\[ e^{\ell(N_c - \frac{N_f}{2} - i N_f \mu \pm i \eta)} \]
Thus the dual partition function is given by:

\[ Z_{N_f,N_f-N_c}^{(U)}(\eta;-m_a;-\tilde{m}_a;-\mu) e^{\ell( N_c - \frac{N_f}{2} - i N_f \mu + i \eta) +\ell( N_c - \frac{N_f}{2} - i N_f \mu - i \eta)} \prod_{a,b} e^{\ell(2 i \mu + i m_a - i \tilde{m}_b)}  \]
Note the extra factors, due to $V_\pm$ and ${M^a}_b$, do not couple to the gauge field and so can be factored out of the integral.  

We wish to show that these two expression are equal for all complex values of the deformations.  One may worry that the partition function does not converge for all values of the deformation parameters.  Indeed, the $1$-loop partition function only decays exponentially, so there is only a finite range of $\mbox{Im}(\eta)$ for which the partition function converges, and similarly for the other parameters.  However, as discussed in \cite{Rains:2010}, there is a natural notion of analytic continuation of a function like this which extends it to a meromorphic function on the space of complex deformations $m_a,\tilde{m}_a,\eta$.  We wish to show the equality of these analytically continued partition functions.

In fact, identities like this one have recently been studied in the mathematical literature \cite{Rains:2010} \cite{vandeBult:2008}.   More precisely, the integrals considered in these papers involved the hyperbolic gamma function $\Gamma_h(z;\omega_1,\omega_2)$, a generalization of the ordinary gamma function which is symmetric in the parameters $\omega_1,\omega_2$, which are fixed and will be suppressed, and which satisfies the following functional equations:

\begin{align}
\label{hgam}
\Gamma_h(z+\omega_1) = 2 \sin (\frac{\pi z}{\omega_2}) \Gamma_h(z) \notag \\
\Gamma_h(z+\omega_2) = 2 \sin (\frac{\pi z}{\omega_1}) \Gamma_h(z) \\ 
\Gamma_h(z) \Gamma_h(\omega_1+\omega_2-z) = 1 \notag
\end{align}
From the first two equations, we see it has an elliptic property that is crucial in proving many of the relevant identities.  As shown in the appendix, this function is related to the $1$-loop partition function by:

\[ \Gamma_h(z;i,i) = e^{\ell(1 + i z)} \]

Actually, taking $\omega_1=\omega_2$ is a somewhat sick case, as the corresponding elliptic curve degenerates, and many of the results need to be checked more carefully in this situation.  However, it was shown in \cite{Hama:2011ea} that if one works on the squashed three sphere, the $1$-loop partition function becomes a double sine function with $b \neq 1$, which corresponds to taking $\omega_1 \neq \omega_2$.  It appears that the formulas above carry over with little modification to this setting, where this problem should not arise, and then the case of an ordinary $S^3$ can be treated as a limiting case.

To see how the identity above follows from the results of these papers, we consider the following integral, defined in \cite{vandeBult:2008}:

\[ I_{n,(2,2)}^m(\mu;\nu;\lambda) = \frac{1}{\sqrt{-\omega_1 \omega_2}^n n!} \int_{C^n} \prod_{1 \leq j<k \leq n} \frac{1}{\Gamma_h(\pm(x_j-x_k))} \prod_{j=1}^n \bigg( e^{\frac{\pi i \lambda x_j}{\omega_1 \omega_2}} \prod_{a=1}^{n+m} \Gamma_h(\mu_a - x_j) \Gamma_h(\nu_a+x_j) dx_j \bigg) \] 
where we define $\Gamma_h(x_\pm) = \Gamma_h(x_+) \Gamma_h(x_-)$.  Here $C$ is a certain contour in the complex plane which we will not define in detail here, except to note that, in the cases relevant for us, it can be taken as the real line.  Using (\ref{hgam}) , one can show that, if we take $\omega_1=\omega_2=i$, we have:

\[ \Gamma_h(\pm z) = (2 \sinh (\pi z) )^{-2} \]
If we also set:

\[ n= N_c, \;\;\; m = N_f - N_c , \;\;\; \mu_a = \frac{i}{2} - \tilde{m}_a + \mu, \;\;\; \nu_a = \frac{i}{2} + m_a + \mu \;\;\; \lambda = -2\eta \]
then one can see that $I_{n,(2,2)}^m$ is precisely the partition function we are studying.  

But now all we need is theorem $5.5.11$ of \cite{vandeBult:2008}, which states:

\[ I_{n,(2,2)}^m(\mu_a;\nu_a;\lambda) = I_{m,(2,2)}^n(\omega-\mu_a;\omega-\nu_a;-\lambda) \prod_{a,b=1}^{n+m} \Gamma_h(\mu_a+\nu_b) \times \]

\[ \times \Gamma_h( (m+1) \omega - \frac{1}{2} \sum_{a=1}^{n+m} (\mu_a+\nu_a) \pm \lambda) c( \lambda \sum_{a=1}^{n+m} (\mu_a-\nu_a)) \]
where $\omega=\frac{1}{2}(\omega_1+\omega_2)$.  If we identify the parameters as above, it's easy to check that the RHS is precisely the partition function of the dual theory.  This demonstrates the partition functions of the two theories are indeed equal.

It may have been unclear in the above calculation what the role of the IR $R$-symmetry was, so let us comment on that now.  The above calculation goes through for complex values of the various mass and FI parameters.  Thus we have actually shown that the partition function for the theories agree even after shifting the $R$ symmetries on both sides by flavor symmetries, provided these symmetries are identified under the duality.  In particular, they must agree for the correct $R$-symmetry, even though, at this point, we do not know what this is.  

Unfortunately, this means the duality cannot be used to find the correct $R$-symmetry.  However, using the discrete symmetries of the two theories, we can constrain the dimensions to have the form:

\begin{equation}
\label{dimss}
\Delta_{Q_a} = \Delta_{\tilde{Q}_a} = \frac{1}{2} + \delta,  \;\;\;\; \Delta_{q_a} = \Delta_{\tilde{q}_a} = \frac{1}{2}-\delta, \;\;\;\; \Delta_{{M^a}_b} = 1 + 2 \delta, \;\;\;\; \Delta_{V_\pm} = \frac{N_f}{2} - N_c + 1 - N_f \delta 
\end{equation}
for some real number $\delta$, which can be identified with the imaginary part of the total axial mass $\mu$.  This means all other deformations may be taken to be real.  We cannot determine $\delta$ at this point, but we will describe an alternative method to determine it later on.

\subsection{Symplectic Group}

Another, similar duality was also studied in \cite{Aharony:1997gp}.  The main difference here is that the gauge group is now symplectic.  The theories are:

\begin{itemize}
\item $\mathcal{N}=2$ $Sp(2N_c)$ gauge theory with $2 N_f$ chiral multiplets $Q_a$ in the fundamental ($2N_c$-dimensional) representation.
\item $\mathcal{N}=2$ $Sp(2(N_f-N_c-1))$ gauge theory with $2N_f$ fundamental chiral multiplets $q_a$.  In addition there are $N_f(2N_f-1)$ uncharged chiral multiplets $M^{ab}$ and a chiral multiplet $Y$, which couple through the superpotential:

\[ M^{ab} q_a q_b + Y \tilde{Y} \]

where, as before, $Y$ and $\tilde{Y}$ parametrize the Coulomb branches of the first and second theories respectively.
\end{itemize}

As in the previous duality, the two theories share the same global symmetries, and they are mapped to each other straightforwardly under the duality, so we may summarize the charges as follows:

\[\]
\begin{center}
\begin{tabular}{| c | c  c c | } 
\hline
Field & $SU(2 N_f)$ & $U(1)_A$ & $U(1)_{R-UV}$ \\
\hline
$Q_a$ & $2N_f$ & $1$ & $\frac{1}{2}$ \\
\hline
$q^a$ & $\bar{2 N_f}$ & $-1$ & $\frac{1}{2}$\\
$M^{ab}$ & $N_f(2N_f-1)$ & $2$ & $1$\\
$Y$ & $1$ & $-2N_f$ & $N_f-2N_c$ \\
\hline
\end{tabular}
\end{center}
\[\]

The contribution of the gauge multiplet is given by (\ref{spnfac}), and the contribution of a chiral multiplet in the fundamental representation, deformed by a mass $m$, is given by:

\[ \prod_{j=1}^{N_c} e^{\ell(\frac{1}{2} + i \lambda_j + i m) + \ell(\frac{1}{2} - i \lambda_j + i m)} \]

Thus the partition function for the first theory, deformed by mass parameters $m_a$, which sum to zero, and axial mass $\mu$, is given by:

\[ Z_{N_f,N_c}^{(Sp)}(m_a) = \frac{1}{N_c!} \int \prod_{j=1}^{N_c} \prod_{a=1}^{2 N_f} e^{\ell(\frac{1}{2} + i \lambda_j + i m_a + i \mu)+\ell(\frac{1}{2} - i \lambda_j + i m_a + i \mu)} \times\]

\[ \times  \prod_{1 \leq i < j \leq N_c} \bigg( (2 \sinh \pi (\lambda_i - \lambda_j))^2 (2 \sinh \pi (\lambda_i + \lambda_j))^2 \bigg) \prod_{j=1}^{N_c} (2 \sinh (2 \pi \lambda_i))^2 \]

For the second theory, the partition function is given by:

\[ Z_{N_f,N_f-N_c-1}^{(Sp)}(-m_a) e^{\ell(2N_c-N_f+1-2 N_f i \mu )} \prod_{1 \leq a<b \leq 2N_f} e^{\ell(i(m_a + m_b + 2 \mu)} \]

In \cite{vandeBult:2008}, integrals of this type were also considered.  Namely, the following definition was made:

\[ I_{n,2}^m (\mu_a) = \frac{1}{\sqrt{-\omega_1 \omega_2}^n n!} \int \prod_{1 \leq j < k \leq n} \frac{1}{\Gamma_h(\pm x_j \pm x_k)} \prod_{j=1}^n \frac{\prod_{a=1}^{2n+2m+2} \Gamma_h(\mu_a \pm x_j)}{ \Gamma_h(\pm 2 x_j)} dx_j \]

Recalling the relations between the $1$-loop partition function and the hyperbolic gamma function discussed above, one can see that this is precisely the partition function of the original theory if we identity:

\[ n=N_c, \;\;\; m = N_f - N_c -1, \;\;\; \mu_a = \frac{i}{2} + m_a + \mu \]

Then theorem 5.5.9 of \cite{vandeBult:2008} says:

\[ I_{n,2}^m(\mu_a) = I_{m,2}^n(\omega-\mu_a) \Gamma_h(2 (m+1)\omega - \sum_{a=1}^{2n+2m+2} \mu_a) \prod_{1 \leq a<b\leq 2n+2m+2} \Gamma_h(\mu_a+\mu_b) \] 

which is precisely the conjectured duality.

\section{Giveon-Kutasov Duality}

Related to the first duality of the previous section is the duality of Giveon and Kutasov.  The main difference is that now there is a Chern-Simons term, and the duality is between groups $U(N_c)$ and $U(|k|+N_f-N_c)$, where $k$ is the Chern-Simons level.  Specifically, the theories are:

\begin{itemize}
\item $\mathcal{N}=2$ $U(N_c)$ gauge theory with $N_f$ flavors and a Chern-Simons term at level $k$.
\item $\mathcal{N}=2$ $U(|k|+N_f-N_c)$ gauge theory with $N_f$ flavors and a Chern-Simons term at level $-k$.  In addition, there are ${N_f}^2$ uncharged chiral multiplets ${M_a}^b$, which couple through a superpotential $\tilde{q}^a {M_a}^b q_b$.  There is no $V_\pm$ field.
\end{itemize}

In \cite{Kapustin:2010mh}, an $\mathcal{N}=3$ version of this duality was considered, which differs from the one here by the addition of an adjoint chiral multiplet and a superpotential coupling the flavors to the vector multiplet.  One nice feature of this version of the duality is that, in flowing to the IR, the only effect is to remove the Yang-Mills term.  Thus we obtain a duality between two superconformal theories for which we can explicitly write down the Lagrangian on both sides.

Returning to the $\mathcal{N}=2$ case, it turns out one can derive this duality from the duality of the previous section as follows.  It is well known that integrating out a massive charged fermion generates a Chern-Simons term at level $\pm \frac{1}{2}$, whose sign is the same as the sign of the mass of the fermion.  Now take a $U(N_c)$ theory with some flavors, and consider adding a large positive mass to one of the flavors.  The flavor can be integrated out, and Chern-Simons terms are generated by each of the two chiral multiplets.  If this is a vector mass, the contributions have opposite signs and cancel, but for an axial mass, they have the same sign, they add up to generate a level one Chern-Simons term.

Let us now consider a general $k>0$.  If we start with a theory with $N_f+k$ flavors and give large positive axial masses to $k$ of the flavors, we generate a level $k$ Chern-Simons term.  This maps to the same operation in the dual theory, albeit with negative axial masses, and so a Chern-Simons term at level $-k$ is generated.  This dual theory has gauge group $U(N_f+k-N_c)$, and this procedure gives a large mass to $V_\pm$ and to some of the $M$ fields, which can then be integrated out.  One can see that we obtain precisely the duality described above.

The considerations above can actually be applied at the level of the matrix model to derive the expected mapping of the partition functions of Giveon-Kutasov duals.  Specifically, we need to look at the asymptotic behavior of the $1$-loop partition function for large mass.  In addition to generating a Chern-Simons term, one finds a constant phase, which one can interpret as being due to the fact that we are computing a Chern-Simons partition function using a non-standard framing of $S^3$, as discussed in \cite{Kapustin:2010mh}.  In fact, a general formula for the mapping of the partition function, including the relative phase, was conjectured in that paper, and we will see that the results here reduce that conjecture to the identity of the partition functions in section $3.1$.

As shown in the appendix, if we take the $1$-loop partition function for a flavor with axial mass $M$, then for $M \rightarrow \pm \infty$:

\[ e^{\ell(\frac{1}{2} + i \lambda +i M) + \ell(\frac{1}{2} - i \lambda + i M)} \approx \exp\bigg(\pm \bigg( - i \pi \lambda^2 - i \pi M^2 - \pi M + \frac{i \pi}{12} \bigg)\bigg) \]
where we have ignored terms exponentially small in $M$.  Note that, up to a $\lambda$-independent factor, this is precisely the contribution to the matrix model of a level-$1$ Chern-Simons term, as expected.

Now consider the partition function $Z_{N_f+1,N_c}^{(U)}(\eta;m_a;\tilde{m}_a;\mu)$, and let the last flavor have a large axial mass $M$, ie, $m_{N_f+1}=-\tilde{m}_{N_f+1}=M$.  We find, for $\mu \rightarrow \infty$:\footnote{Here we impose $\sum_{a=1}^{N_f} m_a =\sum_{a=1}^{N_f} m_a = 0$, ie, we do not include $m_{N_f+1}$ in the sum.  In addition, we ignore $\mu$ relative to $M$ in the last flavor (alternatively, we can absorb it into the definition of $M$).}

\[ Z_{N_f+1,N_c}^{(U)}(\eta;m_1,...,m_{N_f},M;\tilde{m}_1,...,\tilde{m}_{N_f},M;\mu) \approx \]

\[ \approx \frac{1}{N_c!} e^{\pm N_c ( - i \pi M^2 - \pi M + \frac{i \pi}{12} )} \int \prod_{j=1}^{N_c} d \lambda_j e^{\mp i \pi {\lambda_j}^2 + 2\pi i \eta \lambda_j} \prod_{a=1}^{N_f} e^{\ell(\frac{1}{2} + i \lambda_j + i m_a + i \mu) + \ell(\frac{1}{2} - i \lambda_j - i \tilde{m}_a + i \mu)} \prod_{i<j} (2 \sinh \pi(\lambda_i-\lambda_j))^2 \]

\[ = (-1)^{N_c(N_c-1)/2} e^{\pm N_c ( - i \pi M^2 - \pi M + \frac{i \pi}{12} )}  Z_{N_f,N_c,k=\pm 1}^{(U)}(\eta;m_a;\tilde{m}_a;\mu) \]
where we have recognized the integral as the partition function for the level $\pm 1$ Chern-Simons matter theory, whose partition function is given, in the general case, by:

\[ Z_{N_f,N_c,k}^{(U)}(\eta;m_a;\tilde{m}_a;\mu) = \;\;\;\;\;\;\;\;\;\;\;\;\;\;\;\;\]

\[ = \frac{1}{N_c!} \int \prod_{j=1}^{N_c} d \lambda_j e^{- k \pi i {\lambda_j}^2 + 2\pi i \eta \lambda_j} \prod_{a=1}^{N_f} e^{\ell(\frac{1}{2}+i\lambda_j+i m_a + i \mu)+\ell(\frac{1}{2}-i\lambda_j+i \tilde{m}_a + i \mu)}\prod_{i \neq j} 2 \sinh \pi(\lambda_i-\lambda_j) \]
Note the difference in sign convention used for theories with a Chern-Simons term, which is due to how we take the product in the $1$-loop determinant for the gauge sector.  Namely, before we made the choice which ensured the $1$-loop determinant was positive, while here we use a convention which is more natural from the group theory perspective (ie, it is simply what one gets by taking a product over all the roots).  It turns out these two choices give the simplest forms of the two types of dualities.

More generally, we find, for $k$ a positive integer and $M \rightarrow \pm \infty$:

\[ Z_{N_f+k,N_c}^{(U)}(\eta;m_1,...,m_{N_f},M,...,M;\tilde{m}_1,...,\tilde{m}_{N_f},M ,...,M;\mu) \approx \]

\[ \approx (-1)^{N_c(N_c-1)/2} e^{\pm k N_c ( - i \pi M^2 - \pi M + \frac{i \pi}{12} )}  Z_{N_f,N_c,\pm k}(\eta;m_a;\tilde{m}_a;\mu) \]
Now we apply the known mapping of partition function from section $3.1$:\footnote{There is a slight subtlety related to the fact that $m_a$ and $\tilde{m}_a$ no longer sum to zero.  It is straightforward to work out how the $1$-loop determinants for ${M^a}_b$ and $V_\pm$ are modified in this case.}

\[ \log (Z_{N_f+k,N_c}^{(U)}(\eta;m_a,M,...,M;\tilde{m}_a,M,...,M;\mu)) = \] 

\[ = \log (Z_{N_f+k,N_f+k-N_c}^{(U)}(\eta;-m_a,-M,...,-M;-\tilde{m}_a,-M,...,-M;\mu)) + \]

\[ +  \ell( N_c - \frac{N_f+k}{2} - i N_f \mu - i k M + i \eta) + \ell( N_c - \frac{N_f+k}{2} - i N_f \mu - i k M - i \eta) + \]

\[ + \sum_{a,b=1}^{N_f} \ell(i(m_a -\tilde{m}_b + 2 \mu)) + k \sum_{a=1}^{N_f} ( \ell(i(m_a + \mu + M)) + \ell(i(M - \tilde{m}_a + \mu)) ) + k^2 \ell(2 i M) \]
If we use the formula above and the asymptotic expansion for $\ell(z)$ described in the appendix, we get, after taking the strict $M \rightarrow \infty $ limit and simplifying:

\[ \log (Z_{N_f,N_c,k}^{(U)}(\eta;m_a;\tilde{m}_a;\mu)) = \log (Z_{N_f,k+N_f-N_c,-k}^{(U)}(\eta;-m_a;-\tilde{m}_a;\mu)) + \]

\[ + \sum_{a,b} \ell(i(m_a - \tilde{m}_b + 2 \mu)) + \frac{\pi i}{12} (k^2 + 3 (k + N_f) (N_f-2) + 2) + \]

\[ + \pi i \eta^2 - \frac{k \pi i}{2} \sum_a ({m_a}^2 + {\tilde{m}_a}^2) + \pi i  N_f(N_f-k) \mu^2 + \pi N_f( k+N_f - 2 N_c) \mu \] 

When we consider the $\mathcal{N}=3$ version of this duality, the only difference is addition of a superpotential and an adjoint chiral of dimension $1$.  These do not affect the matrix model, but the extended supersymmetry means one cannot allow axial masses, so we must set $m_a = \tilde{m}_a$ as well as $\mu=0$.  In this case, the above formula reduces to:

\[ \log (Z_{N_f,N_c,k}^{(U)}(\eta;m_a)) = \log (Z_{N_f,k+N_f-N_c,-k}^{(U)}(\eta;-m_a)) + \]

\[ + \frac{\pi i}{12} (k^2 + 3 (k + N_f) (N_f-2) + 2) + \pi i \eta^2 - k \pi i \sum_a {m_a}^2\]
This agrees with the the results of \cite{Kapustin:2010mh}, where it was proved in the cases $N_f=0,1$, but only conjectured for larger $N_f$.  Although we have only considered the case where $k>0$ in the original theory, since the dual theory has $k<0$, it is straightforward to invert these formulas to obtain the duality in the case where the original theory has $k<0$.

\subsection{Symplectic Case}

Although Giveon and Kutasov only considered unitary gauge groups, the argument above is easily adapted to the symplectic case.  Consider an $Sp(2N_c)$ gauge theory with $2(N_f+k)$ chiral multiplets.  Now we let the masses for $2k$ of the chiral multiplets be $M$, which we send to $\pm \infty$.  Then we find:

\[ Z_{N_f+k,N_c}^{(Sp)}(\eta;m_1,...,m_{2 N_f},M,...,M;\mu) \approx \]

\[ \approx \frac{e^{\pm 2 k N_c(- i \pi M^2 - \pi M + \frac{i \pi}{12})}}{N_c!} \int \prod_{j=1}^{N_c} e^{\mp 2 k \pi i {\lambda_j}^2} \prod_{a=1}^{2 N_f} e^{\ell(\frac{1}{2} + i \lambda_j + i m_a + i \mu)+\ell(\frac{1}{2} - i \lambda_j + i m_a + i \mu)} \times\]

\[ \times  \prod_{1 \leq i < j \leq N_c} \bigg( (2 \sinh \pi (\lambda_i - \lambda_j))^2 (2 \sinh \pi (\lambda_i + \lambda_j))^2 \bigg) \prod_{j=1}^{N_c} (2 \sinh (2 \pi \lambda_i))^2 \]

\[ = (-1)^{N_c} e^{\pm 2 k N_c(- i \pi M^2 - \pi M + \frac{i \pi}{12})} Z_{N_f+k,N_c,\pm k}^{(Sp)}(\eta;m_1,...,m_{2 N_f};\mu) \]
where we have defined the partition function for a Chern-Simons matter theory with symplectic gauge group by:

\[ Z_{N_f,N_c,k}^{(Sp)}(\eta;m_1,...,m_{2 N_f};\mu) = (-1)^{N_c} \int \prod_{j=1}^{N_c} e^{- 2 k \pi i {\lambda_j}^2} \prod_{a=1}^{2 N_f} e^{\ell(\frac{1}{2} + i \lambda_j + i m_a + i \mu)+\ell(\frac{1}{2} - i \lambda_j + i m_a + i \mu)} \times\]

\[ \times  \prod_{i<j} \bigg( (2 \sinh \pi (\lambda_i - \lambda_j))^2 (2 \sinh \pi (\lambda_i + \lambda_j))^2 \bigg) \prod_{j=1}^{N_c} (2 \sinh (2 \pi \lambda_i)) \]
As before, we use the natural sign convention in the $1$-loop gauge determinant when dealing with Chern-Simons theories.  Also, there is an extra factor of $2$ in the Chern-Simons contribution relative to the unitary case, which is due to the normalization of the generators for the Lie algebra.  One can check this by making sure $SU(2)$ and $Sp(2)$ give the same contribution.

Applying the duality to this theory, we find:

\[ \log Z_{N_f+k,N_c}^{(Sp)}(\eta;m_a,M,...,M;\mu) =  \log Z_{N_f+k,N_f+k-N_c-1}^{(Sp)}(-m_a,-M,...,-M;-\mu) + \]

\[ + \ell(2N_c-N_f-k+1-2 N_f i \mu - 2 i k M )+ \sum_{1 \leq a<b \leq 2N_f} \ell(i(m_a + m_b + 2 \mu)) + 2 k \sum_{a=1}^{2N_f} \ell(i( m_a + \mu + M)) + k(2k-1) \ell(2 i M) \]
Taking the limit $M \rightarrow \infty$ as before, we obtain:

\[ \log Z_{N_f+k,N_c,k}^{(Sp)}(\eta;m_a;\mu) =  \log Z_{N_f+k,N_f+k-N_c-1,-k}^{(Sp)}(-m_a;-\mu) + \]

\[ + \sum_{a<b} \ell(i (m_a + m_b + 2 \mu)) - k \pi i \sum_a {m_a}^2 + 2 N_f(k+N_f) \mu^2 + 2 N_f (2 N_c - N_f - k +1 ) i \mu \]

\[ + \frac{\pi i }{12} ( 2 k^2 + 6 N_f (k + N_f +2) + 15 k + 7 ) \]

In fact, all the formulas above make sense even if $N_f$ and $k$ are half-integral, provided that their sum is an integer so that $2(N_f+k)$ is even.  Thus there are dualities involving theories with an odd number of chiral multiplets (recall that there are $2N_f$ such multiplets) as long as we include a half-integral Chern-Simons term.  Thus we are led to propose a duality between the following theories:

\begin{itemize}
\item $\mathcal{N}=2$ $Sp(2 N_c)$ gauge theory with $2 N_f$ chiral multiplets $Q^a$ and a Chern-Simons term at level $k$.  Here $k$ and $N_f$ may be half-integral, but must sum to an integer.
\item $\mathcal{N}=2$ $Sp(2(|k|+N_f-N_c-1))$ gauge theory with $2 N_f$ chiral multiplets $q^a$ and a Chern-Simons term at level $-k$.  In addition, there are $N_f(2N_f-1)$ uncharged chiral multiplets $M_{a b}$, which couple through a superpotential $M_{ab} q^a q^b$.
\end{itemize}

\section{Dimension by $|Z|$ Extremization}

So far we have been able to provide evidence for the equivalence of the IR fixed points of several $\mathcal{N}=2$ theories.  In these theories, the fields generically have anomalous conformal dimension, and we were able to provide this evidence despite the fact that we did not know what the correct IR dimension was.  As described earlier, this was made possible by the fact that different $R$-symmetries differ by flavor symmetries, and since we know how these map between the dual theories, we can match the partition functions for any possible choice of $R$-symmetry.  Nevertheless, it would be interesting to know which of these is the correct choice.  As argued in \cite{Jafferis:2010un}, this choice should be picked out by extremizing $|Z|$.

Let us briefly comment on that problem now.  Consider the theories of section $3.1$, namely, $\mathcal{N}=2$ $U(N_c)$ gauge theories with $N_f$ flavors.  As argued above, the $R$-symmetry may only mix with the $U(1)_A$ current, and so, in terms of the partition function, we only need to consider giving an imaginary part to the total axial mass $\mu$.  Let us assume there are no mass or FI deformations, so that the real part of all the mass terms are zero.  Then the partition function can be written as a function of the imaginary part of $\mu$, which we'll call $\delta$:

\begin{align}
\label{Zdelt}
Z_{N_f,N_c}^{(U)}(\delta) = \frac{1}{N_c!} \int \prod_j d \lambda_j e^{N_f \ell(\frac{1}{2} - \delta + i \lambda_j)+N_f \ell(\frac{1}{2} - \delta - i \lambda_j)} \prod_{i<j} (2 \sinh \pi (\lambda_i - \lambda_j))^2 \end{align}

According to \cite{Jafferis:2010un}, the physical value of $\Delta_Q$ is determined by extremizing $|Z_{N_f,N_c}^{(U)}(\delta)|$ with respect to $\delta$.  This expression is equal to the corresponding expression for the dual theory:

\[ e^{2 \ell( N_c - \frac{N_f}{2} + N_f \delta) - {N_f}^2 \ell(2 \delta)}  Z_{N_f,N_f-N_c}^{(U)}(-\delta)  \]

One can determine the extremal value of $\delta$ using either expression.  The dimensions of the various fields are then given in terms of $\delta$ by:

\[ \Delta_Q = \frac{1}{2} + \delta, \;\;\; \Delta_q = \frac{1}{2} - \delta,\;\;\; \Delta_M = 1 + 2 \delta, \;\;\; \Delta_V = \frac{N_f}{2} - N_c + 1 - N_f \delta \]

Let us see how this works in a few examples.  First we consider theories with $N_f=N_c=N$, for which the dual theories have no gauge group.  As shown in \cite{Aharony:1997bx}, there is an alternative description with the same matter content as the dual theory, but with the superpotential replaced by:

\[ W = - V_+ V_- \det M \]

For $N=1$, we have a theory of three chiral fields interacting via a cubic superpotential, namely the $XYZ$ theory, and as shown in \cite{Jafferis:2010un}, the partition function function is extremized by setting all fields to have dimension $\frac{2}{3}$.  For $N=2$, the superpotential is marginal, but as we will see in a moment, the extremization argument suggests that the theory is free in the IR, so that the superpotential must be marginally irrelevant.  For $N>2$, the superpotential is irrelevant, and so we expect the theory to be free in the IR.

To see if this follows from the extremization method, note that, in the case $N_f=N_c=N$ there is no integral in the dual partition function, so the duality provides an evaluation formula for the integral:

\[ Z_{N,N}^{(U)}(\delta) =  e^{2 \ell( N (\frac{1}{2} + \delta)) - N^2 \ell(2 \delta)} \]

This expression is real and positive, so we may extremize it by extremizing its logarithm (using $\frac{d\ell}{dz} = - \pi z \cot (\pi z)$):

\begin{align}
\label{ext}
0 = \frac{d}{d \delta} Z_{N,N}^{(U)}(\delta) = \frac{d}{d\delta} (2 \ell( N (\frac{1}{2} + \delta)) - N^2 \ell(2 \delta))\notag \\ 
= - 2 N^2 \pi \bigg( (\frac{1}{2} + \delta) \cot \pi N (\frac{1}{2} + \delta) -2 \delta \cot (2 \pi \delta) \bigg) 
\end{align}

In general, this is a transcendental equation with irrational solutions.  There are a few exceptions.  For example, $N=1$ has $\delta=-\frac{1}{6}$ as a solution, corresponding to the known result $\Delta_M=\Delta_V=\frac{2}{3}$, and $N=2$ has $\delta=-\frac{1}{4}$, corresponding to $\Delta_M = \Delta_V=\frac{1}{2}$, ie, the dual theory is free.

For $N>2$, we can see it is impossible to make the dual theory free, since $\delta$ must be $-\frac{1}{4}$ for $M$ to be free, which then fixes $\Delta_V = 1 - \frac{N}{4} \neq \frac{1}{2}$.  In these theories, there must be hidden symmetries coming from the free fields which appear only in the IR, and these provide the extra freedom which allows us set the dimensions of both $M$ and $V$ to $\frac{1}{2}$.  It is not clear what these extra symmetries map to under the duality.  Curiously, $\delta=-\frac{1}{4}$ is still a (non-unique) solution to (\ref{ext}) whenever $N=2 \;(\mbox{mod } 4)$, although it is not clear what, if any, signficance this has.

One can apply a similar argument to the theories of section $4$ with $|k|+N_f=N_c$, for which the second theory again has no gauge group.  Now there are no $V_\pm$ fields, and it is straightforward to show that taking $\Delta_M=\frac{1}{2}$ is always possible, and gives an extremum.

In cases with $N_f > N_c$, there does not appear to be an evaluation formula for the integral defining the partition function, and we are forced to try find the extrema numerically.  In table $1$ we collect a few results for small $N_f$ and $N_c$, and we also allow a non-zero Chern-Simons term $k$.  They appear to approach $1/2$ from below as $N_c/(k+N_f)$ decreases.  Note that the unitarity bound is $1/4$, since otherwise the gauge invariant chiral primary $Q_a \tilde{Q}_b$ has dimension less than $1/2$, and there is at least one theory here, $U(3), N_f=3$, which violates this bound.

\begin{table}
\begin{center}
\begin{tabular}{| c | c c c c | } 
\hline
$k=0$ & $N_f=1$ & $N_f=2$ & $N_f=3$ & $N_f=4$ \\
\hline
$N_c=1$ & $1/3$ &  $0.4085$ & $0.4369$ & $0.4519$  \\
$N_c=2$ & - & $1/4$ & $0.3417$ & $0.3852$ \\
$N_c=3$ & - & - & $0.2181$ & $0.3058$ \\
\hline
\end{tabular}
\[\]
\begin{tabular}{| c | c c c c | } 
\hline
$k=1$ & $N_f=1$ & $N_f=2$ & $N_f=3$ & $N_f=4$  \\
\hline
$N_c=1$ & $0.4084$ & $0.4198$ & $0.4407$ & $0.4535$ \\
$N_c=2$ & $1/4$ & $0.3107$ & $0.3591$ & $0.3914$ \\
$N_c=3$ & - & $1/4$ & $0.2878$ & $0.3278$\\
\hline
\end{tabular}
\[\]
\begin{tabular}{| c | c c c c | } 
\hline
$k=2$ & $N_f=1$ & $N_f=2$ & $N_f=3$ & $N_f=4$  \\
\hline
$N_c=1$ & $0.4256$ & $0.4368$ & $0.4482$ & $0.4572$ \\
$N_c=2$ & $0.3559$ & $0.3618$ & $0.3838$ & $0.4037$ \\
$N_c=3$ & $1/4$ & $0.3016$ & $0.3284$ & $0.3528$ \\
\hline
\end{tabular}
\end{center}
\caption{Values of $\Delta_Q$ which extremize (specifically, minimize) $|Z|$ for some small values of $N_c,N_f,k$, obtained numerically.  Besides $N_c=N_f=1,2$ and $N_c=k+N_f$, none of these values appear to be rational.}
\end{table}

We close this section with a point about convergence of the partition functions.  As shown in the appendix, the $1$-loop partition function has exponential behavior for large $\lambda_j$.  For theories with a Chern-Simons term, one can add a small imaginary part to $k$, and the gaussian term will dominate this exponential behavior, so the partition function always converges.  However, if there is no Chern-Simons term, a straightforward calculation shows that convergence of the partition function requires:

\[ \Delta_Q < \frac{N_f-N_c+1}{N_f} \]
Although one can define the partition function outside of this range by analytic continuation, one might hope that for physical values of the dimension (ie, those determined by extremization of $|Z|$), this is not necessary.

However, this cannot be the case.  If one takes $N_c/N_f \rightarrow 0$, it can be shown using the large $N_f$ approximation that $\Delta_Q \rightarrow \frac{1}{2}$.  But in the dual theory, where $N_c/N_f \rightarrow 1$, this implies $\Delta_q =1-\Delta_Q\rightarrow \frac{1}{2}$, which is outside the range of convergence.  Thus one is forced to define the partition function by analytic continuation in at least some cases.  In fact, inspecting the table above, we can see that already for $N_c=1, N_f=4$, the dimension we obtained by numerical extremization is outside the range of convergence for the dual $N_c=3, N_f=4$ theory.

\section{Conclusion}

In this paper we studied dualities between $\mathcal{N}=2$ theories in three dimensions reminiscent of Seiberg duality.  We showed that the equality of the partition functions of these theories was equivalent to certain recently discovered integral identities involving the hyperbolic gamma function.  We also discussed how to obtain dualities involving Chern-Simons terms from these dualities by integrating out flavors, and demonstrated the matching of their partition functions.

One might wonder if we can obtain a deeper understanding of these dualities by studying how these mathematical identities are proven.  In many cases, these identities are proven in a similar way to the method used in the current paper to derive Giveon-Kutasov dualities.  Namely, one starts with a known duality and takes certain parameters to infinity, recovering the duality you are interested in.  It is likely this kind of argument can be repeated directly in the field theory description, much like it was for the Giveon-Kutasov theories.  In this way one can reduce the entire class of dualities to some much smaller class.  Going the other way, it is also likely one can obtain new dualities by performing these kinds of manipulations.

In addition, we looked at some implications of the proposal of that the correct IR $R$-symmetry is determined by extremization of the partition function.  We found that it is likely that the partition function must be defined by analytic continuation in some cases, and in others there may be hidden symmetries which restrict the applicability of this method.  Nevertheless, we were able to recover the fact that a certain class of theories were free in the IR using this method.

\appendix

\section{Appendix} 

\subsection{Properties of $1$-loop Partition Function}

In \cite{Jafferis:2010un} the following function was considered:

\[ \ell(z) = - z \log ( 1 - e^{2 \pi i z} ) + \frac{i}{2} ( \pi z^2 + \frac{1}{\pi} \mbox{Li}_2(e^{2 \pi i z})) - \frac{i \pi}{12} \]
It's defining property is:

\[ \frac{d \ell}{dz} = - \pi z \cot ( \pi z) \]
along with $\ell(0)=0$, which means it is odd in $z$.  We will be interested in the function:

\[ f(x) = e^{\ell(\frac{1}{2} + i x)} \]
which is basically the $1$-loop partition function of a chiral multiplet.  It is straightforward to show that it satisfies:

\begin{align}
\label{funceqn}
f(x) f(i - x) = 1 \\
f(x) f(-x) = \frac{1}{2 \cosh (\pi x)} \notag
\end{align}

Recall the functional equation satisfied by the hyperbolic gamma function $\Gamma_h(z;\omega_1,\omega_2)$:

\begin{align}
\Gamma_h(z+\omega_1) = 2 \sin (\frac{\pi z}{\omega_2}) \Gamma_h(z) \notag \\
\Gamma_h(z+\omega_2) = 2 \sin (\frac{\pi z}{\omega_1}) \Gamma_h(z) \\ 
\Gamma_h(z) \Gamma_h(\omega_1+\omega_2-z) = 1 \notag
\end{align}
One can check that $\Gamma_h( \frac{i}{2} + x; i,i)$ satisfies the same functional equations as $f(x)$.  In fact, by relating both functions to the double sine function, as was done for the former in \cite{vandeBult:2008} and the latter in \cite{Hama:2010av}, one can rigorously show they are equal.

Next we consider the asymptotic properties of this function.  For $\mbox{Im}(z)$ large and positive, one can see that:

\[ \ell(z) \approx  \frac{i\pi}{2}  z^2 - \frac{i \pi}{12} \]
up to terms exponentially small in $\mbox{Im}(z)$.  Since $\ell$ is odd in $z$, we have, for $\mbox{Im}(z)$ large and negative:

\[ \ell(z)  \approx - \frac{i\pi}{2}  z^2 + \frac{i \pi}{12} \]

Now consider the $1$-loop determinant of a hypermultiplet, whose chiral multiplets have the same mass $\mu$ (called an axial mass):

\[ Z(\lambda;\mu) = f(\lambda + \mu) f(-\lambda + \mu) \]
Then, using the limits of $\ell(z)$ above, one finds that, for $|\lambda| >> |\mu|$:

\[ Z(\lambda;\mu) \approx e^{- 2 \pi |\lambda| ( \frac{1}{2} + i \mu)} \]
This shows that the asymptotic behavior of the integrands of the partition functions of section $3$ is exponential, and they only converge for a finite range of $\mbox{Im}(\mu)$.

Another limit of interest $\mu \rightarrow \pm \infty$, in which case we find:

\[ Z(\lambda;\mu) \approx e^{\pm  (- i \pi \lambda^2 - i \pi \mu^2 - \pi \mu + \frac{i \pi}{12} )} \]
This confirms that a Chern-Simons term is generated when we integrate out a fermion by giving it a large axial mass.

\bibliographystyle{jhep}

\end{document}